\documentclass[10pt, a4paper, twocolumn]{article}

\usepackage[english]{babel}
\usepackage{microtype}
\usepackage{amsmath,amsfonts,amsthm}
\usepackage[svgnames]{xcolor}
\definecolor{AmericanRose}{rgb}{1.0, 0.01, 0.24}
\definecolor{CadmiumRed}{rgb}{0.89, 0.0, 0.13}
\definecolor{Burgundy}{rgb}{0.5, 0.0, 0.13}
\definecolor{DarkGoldenrod}{rgb}{0.72, 0.53, 0.04}
 	
\usepackage[small, labelfont={color=AmericanRose,bf}, up, textfont=it]{caption}

\usepackage{booktabs}
\usepackage{float}
\usepackage{lastpage}
\usepackage{graphicx}
\usepackage{tikz}
\usetikzlibrary{backgrounds}
\usepackage{enumitem}

\newsavebox{\picbox}

\setlist{noitemsep}
\usepackage[framemethod=tikz]{mdframed}

\usepackage{geometry}
\usepackage[T1]{fontenc}
\usepackage[utf8]{inputenc}

\usepackage{XCharter}

\usepackage{fancyhdr}
\newcommand{\ignore}[1]{}

\fancypagestyle{firstpage}{ %
	\fancyhf{}
}

\newcommand{\authorstyle}[1]{{\large\usefont{OT1}{phv}{b}{n}#1}} %

\newcommand{\institution}[1]{{\footnotesize\usefont{OT1}{phv}{m}{sl}\color{Black}#1}} %

\usepackage{titling} %

\newcommand{\HorRule}{\color{DarkGoldenrod}\rule{\linewidth}{1pt}} %
\newcommand{\HorRuleRed}{\color{DarkRed}\rule{\linewidth}{1pt}} %

\pretitle{
	\vspace{-30pt}
	\HorRule\vspace{10pt}
	\fontsize{18}{24}\usefont{OT1}{phv}{b}{n}\selectfont
	\color{DarkRed} %
}

\posttitle{\par\vskip 15pt}

\preauthor{}

\postauthor{
	\vspace{5pt}
	\par\HorRule
	\vspace*{-27pt}
}

\usepackage{aurical}

\usepackage{lettrine}
\usepackage{fix-cm}

\newcommand{\initial}[1]{ %
	\lettrine[lines=5,findent=16pt,nindent=0pt]{%
		\color{DarkRed}%
		{#1}%
	}{}%
}

\newcommand*\circled[1]{\tikz[baseline=(char.base)]{
            \footnotesize\node[shape=circle,draw,inner sep=1pt] (char) {#1};}}

\usepackage{xstring}

\mdfdefinestyle{RedFrame}{
    outerlinewidth=2pt,
    roundcorner=5pt,
    innertopmargin=\baselineskip,
    innerbottommargin=\baselineskip,
    innerrightmargin=10pt,
    innerleftmargin=10pt,
    linecolor=DarkRed,
    leftmargin=-0.4cm,
    rightmargin=-0.4cm,
    }

\usepackage[backend=biber,style=ieee,natbib=true,doi=false,isbn=false,url=false,eprint=false]{biblatex}

\usepackage[autostyle=true]{csquotes}

\usepackage{wrapfig}

\usepackage{bbding}

\usepackage{caption}
\usepackage[hidelinks,destlabel=true]{hyperref} 
\usepackage{pdfpages}

\usepackage{eso-pic}
\usepackage{color}
\definecolor{brown}{rgb}{0.5,0.3,0}
\definecolor{ruby}{rgb}{0.6,0,0.3}
\definecolor{maroon}{rgb}{0.8,0,0.4}
\definecolor{rose}{rgb}{1.,0,0.4}
\definecolor{orange}{rgb}{1.0,0.4,0.0}
\definecolor{emerald}{rgb}{0.0,0.5,0.0}
\definecolor{myblue}{rgb}{0.1,0.1,1}
\definecolor{linkblue}{rgb}{0,0.4,0.8}

\newcommand{\link}[1]{{\color{linkblue}#1}}

\graphicspath{{./}}

\title{\protect\centering Automation and AI Technology in Surface Mining\\[1mm]\normalsize With a Brief Introduction to Open-Pit Operations in the Pilbara$^\ddag$}

\author{
    {
        \protect\centering \protect\includegraphics[width=0.78\linewidth,trim={0 -3mm 0 0},clip]{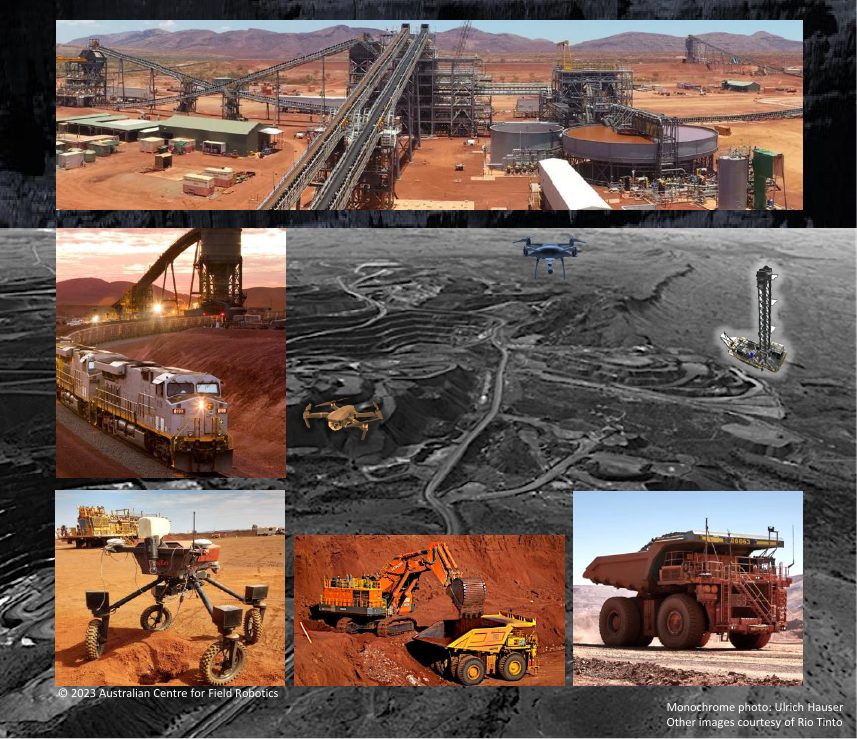}\\
        \authorstyle{By Raymond Leung, Andrew J Hill and Arman Melkumyan}\\[1mm]
    }
    {\institution{\hspace{35mm}\,Australian Centre for Field Robotics, The University of Sydney, NSW 2006 Australia}}
}

\date{}

\begin{document}

\AddToShipoutPicture*{
     \AtTextUpperLeft{
         \put(-3,11){
           \begin{minipage}{\textwidth}
              \textit{\,}
           \end{minipage}}
     }
}

\maketitle

\thispagestyle{fancy}
\renewcommand*{\thefootnote}{\fnsymbol{footnote}}

\textbf{Abstract}---This survey article provides a synopsis on some of the engineering problems, technological innovations, robotic development and automation efforts encountered in the mining industry---particularly in the Pilbara iron-ore region of Western Australia. The goal is to paint the technology landscape and highlight issues relevant to an engineering audience to raise awareness of AI and automation trends in mining. It assumes the reader has no prior knowledge of mining and builds context gradually through focused discussion and short summaries of common open-pit mining operations. The principal activities that take place may be categorized in terms of resource development, mine-, rail- and port operations. From mineral exploration to ore shipment, there are roughly nine steps in between. These include: geological assessment, mine planning and development, production drilling and assaying, blasting and excavation, transportation of ore and waste, crush and screen, stockpile and load-out, rail network distribution, and ore-car dumping. The objective is to describe these processes and provide insights on some of the challenges\,/\,opportunities from the perspective of a decade-long industry-university R\&D partnership.\footnote[3]{\scriptsize\textcopyright 2023 IEEE.  Personal use of this material is permitted.  Permission from IEEE must be obtained for all other uses, in any current or future media, including reprinting/republishing this material for advertising or promotional purposes, creating new collective works, for resale or redistribution to servers or lists, or reuse of any copyrighted component of this work in other works.}
\renewcommand*{\thefootnote}{\arabic{footnote}}

{
\noindent\HorRuleRed
\\[0.2cm]
\color{DarkRed}
\textbf{(Picture) Mining automation encompasses mine, rail and port operations. Robotic platforms and data analytics are being used increasingly in high-tech mines \cite{rio-innovation2021}. Examples include autonomous drills, haul trucks, shovels, conveyors, drones, trains and ships.}
\\
\HorRuleRed
}

\newpage
\section*{\color{AmericanRose}Foreword}
\initial{I}{n} the last decade, there has been a concerted effort aimed at improving mining efficiency and safety through mining automation. The benefits of increased productivity and reduced exposure to occupational hazards for workers at open-pit mines are easy to appreciate. What is less obvious is understanding the type of engineering problems surface mining operations pose, and how interdisciplinary research or innovative ideas from other fields can be brought to bear on these problems. This article attempts to explain these issues to a general audience in an accessible way. It lays bare some of the engineering problems, technological innovations, robotic development and automation efforts encountered in the mining industry. In so doing, it broadens current perspective of what automation, robotics and intelligent systems can offer to industries, particularly the surface mining industry. Examples and references are cited to illustrate how machine learning, remote sensing, geostatistics, planning and optimization are used to address specific problems within such large-scale interconnected systems. Readers can expect to develop an understanding of the key processes involved, identify some of the challenges and opportunities in mining and potential areas they may contribute to in the future.

A distinctive feature about this survey article is that its contents are layered and varied. While it focuses on engineering problems and innovations in surface mining, it also provides insights on state-of-the-art technologies and future trends. This discussion is informed by multidisciplinary perspectives and experiences from working in different fields.

\section*{\color{CadmiumRed}Overview}
At the time of writing, it has been sixteen years since Durrant-Whyte, Nettleton, Nebot, Scheding et al. articulated the vision of an autonomous, remotely-operated mine through a series of publications \cite{nebot2007surface,vasudevan2010mine,marshall2016robotics,madhavan1998map,roberts2000autonomous,le1998towards,widzyk2008millimetre,elinas2009multi,fan2010integrated,thompson2011distributed,nieto20103d} and patents \cite{durrant2012method,nettleton2014method,nettleton2015control,elinas2015drill,mchugh2019method,nettleton2016integrated,silversides2011patent,silversides2013patent,rtgi-rtcma2018,hill2021patent,rtgi-rtcma-rail2021,vujanic2022patent}. These seminal works and engineering contributions have left a lasting legacy and propelled the Australian mining industry to the forefront in terms of technology investment and adoption \cite{rio-ami2017,rio-incubator-automation2019}. Unmanned Aerial Vehicles (UAV) \cite{shahmoradi2020comprehensive,equinox2022,hovermap2022} and remote sensing technologies are being used in surveys to boost productivity and improve mine planning, safety and inventory management. Other significant examples and historical outcomes include the deployment of autonomous drills, opening of the world-first remote mining operation center based in Perth, and running fully autonomous trains for transporting iron ore from pit to port across the Pilbara. Part of our motivations in writing this paper are to provide an update on more recent developments, highlight the interdisciplinary nature of this work and acknowledge the contributions of our colleagues, both past and present.

In the early days (c.\,1998--2006), project proposals focused intently on mapping, localization and navigation problems \cite{durrant2006simultaneous,bailey2006simultaneous,nieto2006scan,perera2009linear,perera2010nonlinear,scheding1999experiment,thrun2004autonomous,seiler2012using} within an unstructured and dynamic mining environment. A key part of the overall vision was (and still is) applying a probabilistic viewpoint to mining; given that it is a large-scale system with many interacting parts, a perfect knowledge spanning geology and equipment operations is nigh impossible. In the intervening years, consolidation of the program has resulted in a sustained period of growth in other research areas: such as Gaussian Processes \cite{vasudevan2009gaussian,vasudevan2010large,melkumyan2009sparse,melkumyan2011multi,melkumyan2011non,jewbali2011apcom}, hyperspectral imaging \cite{murphy2012evaluating,murphy2013mapping,murphy2014gaussian,murphy2015mapping,schneider2014evaluating,ramakrishnan2015hyperspectral}, geostatistics and boundary modeling \cite{chlingaryan2014patent,leung2019sample,balamurali2019comparison,mery2017geostatistical,leung2019subsurface,leung2021bayesian,silversides2021boundary,ball2022creating,ball2021geologist}, geomechanics and rock recognition \cite{leung2015automated,monteiro2009conditional,zhou2009spectral,zhou2010automated}, production scheduling and truck dispatch \cite{samavati2018new,samavati2019improvements,samavati2020production,seiler2020flow,blom2018multi}, system control \cite{zubizarreta2011multi,maeda2012learning,maeda2014iterative,maeda2015combined,wallace2021experimental,innes2011estimation}, sensor calibration \cite{underwood2010error,taylor2012mutual,taylor2013automatic,taylor2016motion}, computer vision\,/\,perception (incl.\,deep learning) \cite{underwood2013explicit,romero2013stereo,romero2013unsupervised,romero2016variational,ramakrishnan2015shadow,windrim2017physics,windrim2018pretraining,windrim2019unsupervised,windrim2023gsf}. Meanwhile, data fusion \cite{durrant2005data,vasudevan2015efficacy}, machine learning/probabilistic reasoning \cite{kadkhodaie2010rock,monteiro2009learning,monteiro2009applying,ahsan-15,jafrasteh2018comparison,leung2022surface,wedge2018data,liu2021ajcai,khushaba2022machine,bailey2022jt} (disciplines deeply rooted in mathematics and AI) and stochastic optimization \cite{palmer2013stochastic,palmer2014stochastic,palmer2017weekly,palmer2018modelling} continue to play a significant role in advancing the field of mining automation. As we delve into processes and operations later in this paper, the bigger picture of interconnectedness will emerge.

{
\noindent\HorRuleRed
\\[0.2cm]
\color{DarkRed}
\textbf{The unifying theme is that an interesting mix of intelligent systems, computational frameworks and hybrid technologies are driving the use and adaptation of automation technology within the mining industry to address specific needs. Increasingly, small and large-scale problems are being solved within integrated workflows.}
\\
\HorRuleRed
}

In terms of impact, the World Economic Forum estimated that digital transformation would generate \$320 billion for the mining and metals sector in the decade leading to 2025 (this equates to 2.7\% of industry revenue and 9\% of industry profit \cite{report-wef2017}) and an estimated value to society and environment of \$30 billion (with a 610 million tonnes reduction in CO$_2$ emissions). A report commissioned by the Intergovernmental Forum on Mining, Minerals, Metals and Sustainable Development also indicated that since the adoption of autonomous haulage systems (AHS technology), operating costs at Rio Tinto have fallen by 13\% (the autonomous fleet outperforms the manned fleet on average by 14\%). Interested readers are referred to \cite{report-wef2017} and \cite{report-igf2018} for more comprehensive analysis, and case studies of technological and business model innovations for major miners around the world. For a survey and opinion pieces on the role of machine learning, AI and process automation in mining and the geosciences, see \cite{dramsch202070} and \cite{ali2020artificial}.

\begin{figure*}[!htb]
\centering
\includegraphics[width=0.95\textwidth,trim={0mm 0mm 0mm 0mm},clip]{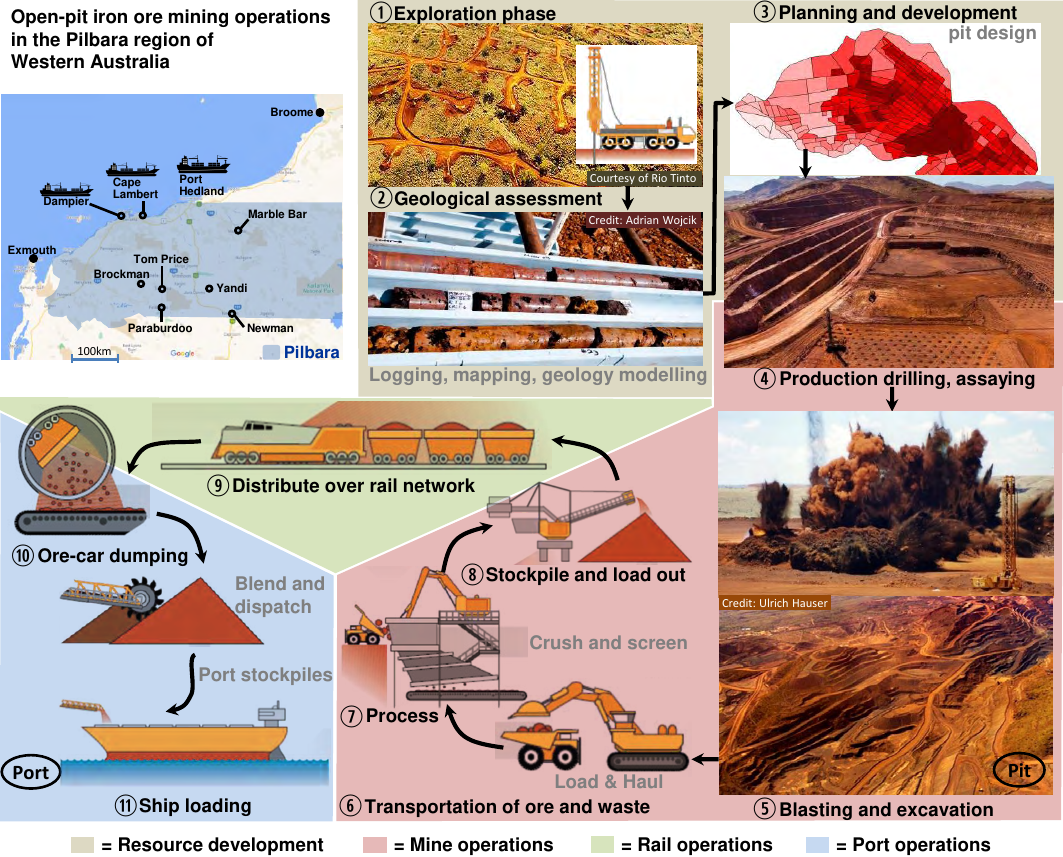}
\caption{Overview of open-pit iron ore mining operations in the Pilbara (adapted from \cite{rt-pilbara-mining-process})} \label{fig:iron-ore-open-pit-mining-pilbara}
\end{figure*}
This article focuses on The Pilbara as it represents a significant part of iron ore mining operations in the southern hemisphere. The Pilbara is an iron-ore mining province located in the northern part of Western Australia, about 400 km east of Exmouth and 320km south of Port Hedland (see Fig.~\ref{fig:iron-ore-open-pit-mining-pilbara}). It encompasses an area of 500,000 km$^2$ and the townships of Brockman, Tom Price, Paraburdoo and Newman, each equipped with its own airport. The indigenous people of Yinhawangka, Banjima, Kurrama and Pinikura (among others) hold native title to this land. During the summer, the temperature regularly exceeds 40$^\circ$C (104$^\circ$F) which makes this environment inhospitable; however, the region still accommodates a FIFO (fly-in fly-out) mining workforce. The main tasks in a typical mining operation involve finding, mining, and processing iron ore. Most of the automation, innovation and research activities relate to geostatistical and geotechnical modeling, mine planning, scheduling, safety, clean energy and quality control. Although these processes and specific practices may vary from site to site, the life cycle of a mine usually begins with resource development (see Fig.~\ref{fig:iron-ore-open-pit-mining-pilbara}). For more information on \href{https://museum.wa.gov.au/research/collections/earth-and-planetary-sciences/rock-collection/banded-iron-formation}{\color{ruby}banded-iron formation} (BIF) hosted iron mineral systems and the geology of (martite-goethite mineralization in) the Hamersley region, please refer to \cite{hagemann2016bif,murphy2017investigating,perring2020new}. The body of this article is organized as follows. The first section (pages~\pageref{sec:processes}-\pageref{sec:engr-research-interests}) provides a background introduction and describes fundamental processes in the mining chain. The second section (pages~\pageref{sec:engr-research-interests}-\pageref{sec:technologies-insights}) briefly examines engineering and research interests in the mining space. It reinforces key concepts and presents a vision of autonomous mining using pictures. The final section (pages~\pageref{sec:technologies-insights}-\pageref{sec:conclusion}) provides insights on traditional and emerging topics that span the mining lifecycle; an example of a mining inspection robot is shown on page~\pageref{fig:bhir}.

\section*{\color{AmericanRose}Processes in the mining chain}\label{sec:processes}
This section aims to familiarize readers with processes in surface mining. The description follows the roadmap\,/\,processes depicted in Fig.~\ref{fig:iron-ore-open-pit-mining-pilbara}. In the exploration phase \circled{1}, geologists attempt to find and determine the viability of ore deposits. A range of geological, geophysical and metallurgical techniques can be used. This generally involves drilling in remote areas and analyzing the samples obtained. Geological assessment \circled{2} is performed whereby the exploration data is logged, mapped, and interpreted through models. Conventional techniques are described in \cite{berg-09,berg-11,berg-15}. Some novel approaches include the use of Dynamic Time Warping with Gaussian Processes (DTW-GP) \cite{silversides2016dynamic,george2021bedding} and hyperspectral imaging \cite{uezato2016novel,uezato2016incorporating,silversides2017identification}. The former identifies stratigraphic boundaries in stratiform deposits using natural gamma logs while the latter estimates mineral abundance.\footnote{In terms of endmember composition, samples may contain variable mixtures of hematite, ochreous and vitreous goethite. Different ratios have different implications for mineral processing.} This preliminary process establishes the location, quantity, and quality of an orebody. Visually, the concentration of a chemical component in an ore deposit is illustrated through various shades of red in Fig.~\ref{fig:iron-ore-open-pit-mining-pilbara}(top-right). During planning and development \circled{3}, engineers determine which orebodies to mine and in what sequence \cite{tabesh2013automatic,mousavi2016open}, to deliver the required product quality at an appropriate cost \cite{rezakhah2020open}. The pit is designed with geotechnical factors (e.g. slope stability \cite{karam2016slope}) in mind to mitigate safety and economic risks, generally the overburden must be removed to access high grade mineral. Mine planning may commence many years before a mine is developed, and it continues on a daily basis once the mine is operational.

Production drilling \circled{4} commences once an area for open-pit mining is selected. Using the mine plan developed, areas are tagged, then blastholes are drilled following some pattern by drill-rigs. Material type logging is performed to assess the proportion of minerals and hardness (or friability) of the material in production blastholes. This assessment provides geometallurgical guidance for blasting and mineral processing in subsequent steps. Importantly, geochemical analysis (assaying) is performed on certain blasthole samples to build knowledge and confidence, since the local geology may deviate from the coarse predictions informed by the resource model. It should be emphasized that the knowledge gained about the orebody from new data may be used to update the geological structure \cite{leung2020mos,leung2021bayesian} (for example, mineralization boundaries) and the uncertainty information provided by a probabilistic model can influence decisions (for example, modify the blast procedure \cite{klerkx2021apcom}/hole spacing in the current/next bench). Holes are loaded and charged with explosives. Blasting fragments the rock in preparation for digging. Excavation \circled{5} and transportation of ore and waste material to various stockpiles or dumps \circled{6} constitute major logistics operations in an open-pit mine. Following blasting, the broken material must be excavated and transported for downstream processing. The term ``load-haul cycles'' is often used to refer to the transfer of material from a pit to the back of trucks using excavators, face shovels or front-end loaders, and moving from a source to another destination. These are heavy machines. Haul trucks can carry a 360t payload, with a kerb weight over 250t. Thus, safe operation is paramount as any collision or incidents can have dire consequences. Mine safety is becoming an increasingly important issue as more automated fleets enter into mining operations. Safety incidents (including near-misses) are reviewed by mining companies and/or investigated by government regulators. The risks to human life are partially offset by having fewer personnel working in potentially hazardous parts of the mine.

The processing of ore \circled{7} ranges from crushing and screening to a standard size, through to processes that beneficiate or upgrade the quality of the iron ore products. This is done by physical processes that remove impurities by differences in particle density, size gravity or size separation in a dry or wet processing plant. Processed ore is stockpiled and blended to meet product quality requirements, before being reclaimed and conveyed to rail load-out \circled{8}. The ore is loaded into ore cars and transported to port facilities by train \circled{9}. In the Pilbara, the rail system is privately owned. It represents one of the most sophisticated rail networks for mining anywhere in the world. Ore is transported over distances of up to 460km to the port. The cumulative distance traveled in the network exceeds 15,000 km per day, which almost equates to a return trip on the Trans-Siberian railway. The driverless train and tracks are controlled remotely by a human operator at a centralized command centre. A freight train typically contains 230 compartments (ore cars) and is 2.4km long. Ore car dumping \circled{10} occurs once the train arrives at the port. Ore cars empty their payload into bins which are then discharged onto conveyors. The port stockpiles are managed according to product type and the quality control plan. Travelling stackers create 250m long stockpiles and rotary bucket-wheel reclaimers later reclaim the ore to load the ships \circled{11}. Although these processes are described as a linear sequence of events, in reality, feedback loops exist between different stages. For instance, Kalman Filter data assimilation approaches have been proposed for material tracking \cite{innes2011estimation} and modeling ore properties \cite{talesh2022real}. In practice, if the material obtained from one location causes handling issues/processing disruption, remedial actions may be taken upstream. Alternatively, suitable high grade material might be sourced from a different location to produce the required blend. Given the numerous variables needed to reflect the spatial, geochemical diversity and temporal planning horizons, this usually results in a complex, large-scale optimization task.

{
\noindent\HorRuleRed
\\[0.2cm]
\color{DarkRed}
\textbf{This complex network and movement of material presents challenges and opportunities with respect to flow planning, devising mining and transportation schedules, and quantification of risks in stochastic optimization.}
\\
\HorRuleRed
}

\section*{\color{CadmiumRed}Research in automation of current mining practices}\label{sec:engr-research-interests}
This section provides a high-level overview of recent research efforts in using automation techniques to improve existing mining practices, and the role automation and digital sciences can play in delivering an autonomous mining future with greater safety, efficiency and productivity. The aim is to provide an understanding and appreciation of the engineering problems in this field, and a foundation for the deeper issues and longer-term challenges discussed in more detail in section~\ref{sec:technologies-insights}.

\begin{figure*}[!htb]
\includegraphics[width=\textwidth,trim={0mm 0mm 0mm 0mm},clip]{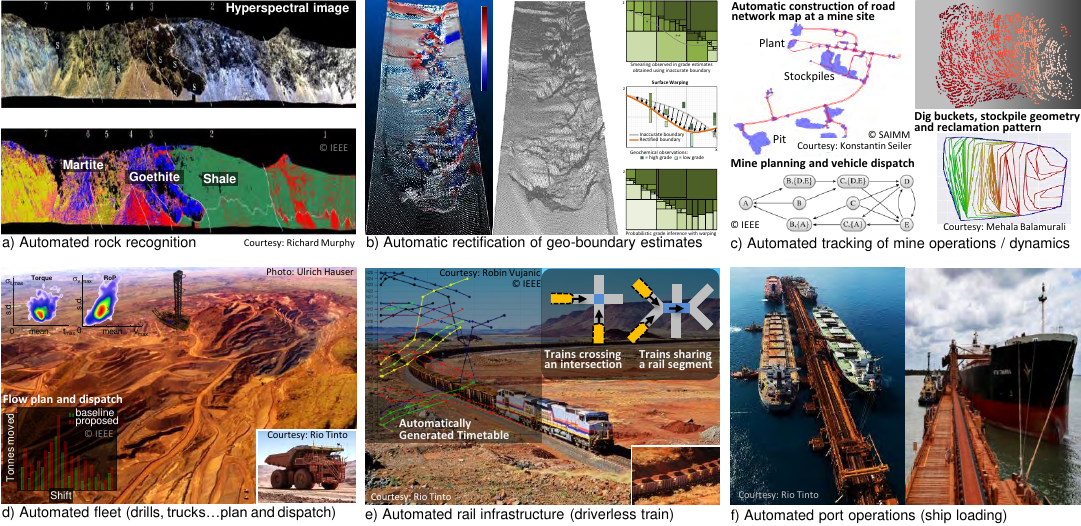}
\caption{\textit{Machines that Sense, Think and Act} --- examples of automation in open-pit mining.} \label{fig:panorama}
\end{figure*}

\newpage
State-of-the-art open-pit mines are becoming increasingly automated. Fig.~\ref{fig:panorama} presents a visual panorama of how automation research is being applied to mining operations in (a) sensing, (b) modeling, (c) tracking via digital twins, and (d--f) planning, dispatch and equipment operation across the various fleets and processes in the mining supply chain.

Novel sensing technologies, autonomous sensing and perception systems are vital for gathering information in an increasingly autonomous operation. Robotic equipment relies on having high quality information about its environment, and mining operations must be constantly informed about the state of the pit or process. Novel sensing technologies can provide expert-level context about the mining environment for modeling and automation. Fig.~\ref{fig:panorama}(a) illustrates the use of hyperspectral image classification to infer the mineralogy from an exposed mine face, as described in \cite{murphy2012evaluating,murphy2013mapping,murphy2014gaussian,murphy2015mapping}.

Automated modeling and tracking converts and fuses gathered information into higher level representations, usable by humans and machines alike. A good model of a system and tracking of its current state is vital to making good control decisions about it.

In geology modeling, mesh surfaces are used extensively to delineate geological domains (such as mineralized and waste zones) where localization errors can significantly impact grade estimation and ore recovery. Fig.~\ref{fig:panorama}(b) shows a surface warping framework \cite{leung2021bayesian} that automatically rectifies the placement of existing boundaries given new geochemical observations. This opportunity arises during the mining production phase as more densely sampled bench assay data becomes available, and automation of this modeling process allows continuously updated orebody models for near-real-time decision-making.

Fig.~\ref{fig:panorama}(c) shows advances in tracking the state of the mine and its assets. Advances in spatiotemporal analysis have allowed accurate maps of vehicle trajectories to be built automatically from vehicle GPS data\cite{seiler2020haulroad}, despite the highly dynamic nature of mining roads and physical constraints. The layout and connectivity of roads is vital to planning material flows and truck dispatch decisions\cite{seiler2020flow,samavati2019improvements}, and creates opportunities for multi-vehicle route optimization \cite{gun2019multi} and improvement to road design or quality for fleet efficiency. This also allows precise tracking of material movement\cite{leung2022material} to measure performance and conformance to plan. The benefits of such tracking applications are increased situational awareness, safety, and conformance in a dynamic environment, independent of interoperability between proprietary systems. By monitoring material movement such as bucket compositional uncertainty \cite{balamurali2022bayesian} and reclamation patterns \cite{balamurali2021apcom-geom} from stockpiles, high-precision tracking can be achieved from the dig face to the process plant and spatial tracking within stockpiles.

With high quality, real-time data, models and state-tracking, it becomes possible to make intelligent decisions in the control of an operation or process. In mining, this is relevant at various operational scales, from equipment to fleet to full supply chain (Fig.~\ref{fig:panorama}(d--f)). While autonomous drills, haul-trucks and trains are the most visible advances in automated operations, the seamless integration of autonomous (and human-operated) equipment is another significant opportunity for process improvement. Fig.~\ref{fig:panorama}(d) shows the increased productivity of modern approaches to flow optimization and haul-truck dispatching \cite{seiler2020flow}, and operating efficiency is similarly enhanced via careful motion planning \cite{gun2019multi}, sequencing \cite{de2017adaptive} and coordination \cite{samavati2019improvements}.

\vspace{5mm}
\begin{figure*}[!bth]
\includegraphics[width=\textwidth,trim={0mm 0mm 0mm 0mm},clip]{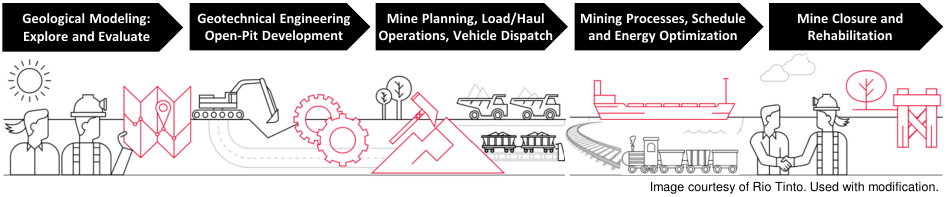}
\caption{Major themes in mining lifecycle} \label{fig:themes-lifecycle}
\end{figure*}
{
\noindent\HorRuleRed
\\[0.2cm]
\color{DarkRed}
\textbf{The ability to handle disruptions (such as refuelling), work around a maintenance schedule \cite{palmer2017weekly}, and respond to changing conditions and unexpected events using real-time information, will be a ubiquitous feature of next generation technology that enables efficient, continuous, autonomous operations.}
\\
\HorRuleRed
}
Various sensing technologies \cite{krolop2019mineralogical,haest2022shovelsense,bona2018das} and hardware~/~software integration with robotic platforms are under development. Applications include mineralogical imaging, equipment health monitoring, predictive modeling for mining performance and ergonomics \cite{ali2018artificial}.

Continuing this line of thinking, with advancements in computational power, optimization algorithms, and connectivity of information systems, there is an opportunity to make globally optimal decisions across the whole supply chain, rather than limiting the decision-horizon to a single vehicle, fleet or operation. Generating feasible train timetables \cite{jones2022trainjournal} to address gridlock and capacity constraints, to efficiently deliver mine stocks to the port via the rail network is highly desirable \cite{vujanic2021computationally}. This means increasing throughput, network utilization and reducing queuing time. There is ongoing interest in linking material movement to the underlying geology, thus acquiring a detailed understanding of how material are sourced and distributed through a network of mines. This has implications for the grade-risk minimization problem, which seeks to find optimal ways to combine the mine stockpiles to produce a consistent blend that stays within the tolerances specified by customers \cite{zhouhn2021heuristics}. The main point to appreciate is that appropriate combination of the stockpiles can lead to lower uncertainty (variability) in the final grade of the exported product.

As engineers, we believe big data analytics, machine learning, operations research, mechatronics and sensing technologies can fundamentally improve efficiency, productivity, robustness, safety in mining. At the same time, the future trajectory is shaped by market forces, thus optimism needs to be measured with due regard for investment appetite and economic conditions. For technically challenging and complex problems, project delivery often takes on a minimum viable product approach \cite{jones2022mvpsim,enkler2019agile} coupled with validation stage gates during a project's life span.

\section*{\color{AmericanRose}Technologies and insights}\label{sec:technologies-insights}
\begin{table*}[t]
\setlength{\tabcolsep}{4pt}
\centering
\resizebox{\textwidth}{!}{
\begin{tabular}{|lp{35mm}|p{70mm}|p{92mm}|}\hline
& Mining themes & Disciplines\,/\,primary concerns & Techniques\,/\,general approaches\\\hline
1 & Geology modeling & Economic geology. Stratigraphy. Geophysics. Geostatistics. Computational geometry. Applied machine learning. Robust statistics. & Kriging + sequential Gaussian simulation. Gaussian processes regression. Ensemble Kalman Filter (EnKF) data assimilation. Geochemistry-based Bayesian deformable surface models. Compositional data analysis.\\\hline
2 & Geotechnical engineering, pit development & Soil/rock mechanics, engineering geology. \nobreak{Decision} making with incomplete information. Probabilistic risk assessment & Pit design for slope stability. Finite element method (FEM). Uncertainty estimation. \nobreak{Computer} graphics: ray-tracing, mesh surface manipulation.\\\hline
3 & Intelligent systems for material tracking and probabilistic inference & Geospatial information system. Probabilistic graphical models. Sequence analysis and state inference. Computer vision and deep learning	& Road network generation using GPS data. Data fusion. Dynamic models. Message passing\,/\,junction trees. Hidden semi-Markov model. Video segmentation/activity recognition.\\\hline
4 & Robotics: sensing and automation & Autonomous agents and human-in-the-loop. Multimodal sensing. Intelligent systems. Reliability engineering in complex systems. Persistent autonomy, interoperability and standards & Hazard reduction/risk minimization in structured problems. Robust design. Adaptive sampling/active learning. Collaborative information gathering. Reinforcement learning. Hyperspectral CNN. OEM innovation.\\\hline
\end{tabular}}
\caption{A quick reference for discussion: Themes--technologies matrix for surface mining in the Pilbara (continues on p.\,\pageref{tab:themes-technologies-continued})}\label{tab:themes-technologies}
\end{table*}

To develop a deeper understanding of the core issues, this section takes a deep-dive into state-of-the-art technologies and provides critical insights on how robotics, automation, AI and ML might be used to tackle current and future challenges in surface mining. This discussion is anchored on several themes which span the mining lifecycle (see Fig.~\ref{fig:themes-lifecycle}). It covers traditional and emerging topics, such as geostatistical modeling, stochastic optimization, carbon-neutral mining and land rehabilitation. For quick reference, these topics are shown in the mining themes--technologies matrix in Table~\ref{tab:themes-technologies}.

\subsection*{Geology (orebody) modeling}
A large part of geology modeling is concerned with inferring the geochemical/metallurgical properties of ore deposits over a dense grid using sparse observations at known locations. In geostatistics, kriging (an optimal linear unbiased estimator) is the method of choice, and this is usually accompanied with sequential Gaussian simulation (SGSim) to compensate for oversmoothing and estimate grade uncertainty. An issue with this is high computation cost. These deficiencies are addressed in a physics-informed machine learning (PIML) proposal \cite{bai2022sequential} where only a small amount of data produced by SGSim is used during training to discover the spatial correlation between the available data and unsampled points. Viewing the target attribute as a stochastic function, Gaussian Processes (GPs) \cite{jewbali2011apcom} offer a non-parametric Bayesian framework for computing the posterior distributions to obtain the predictive mean and covariance (uncertainty).\footnote{The connections between kriging and GPs are described in \cite{shekaramiz2019note}.} The basic premise is that spatial structures can be described by a covariance matrix \cite{melkumyan2009sparse}. For mining, the formulation of covariance functions (kernels) is modified to cater for volumetric data fusion which allows blasthole assays to be considered as interval or cross-zone measurements rather than points \cite{chlingaryan2014patent}. This enables fast, accurate block-based ore property estimation that computes for the block dimensions without resorting to point-based approximation.

Recent advances have focused on model refinement using incremental data. The common theme here are approaches for updating geological boundaries using blasthole data as bench-mining progresses. The main point to appreciate is that geochemical distributions are neither isotropic nor homogeneous. Na\"{i}ve interpolation across geological discontinuities would produce smearing artifacts across distinctive layers \cite{silversides2017identification}. Hence, unbiased grade prediction depends critically on having accurate geological surfaces (e.g. mineralization boundaries) that delineate the modeling space by regions and annotate observations according to geological domains. From a control-optimization perspective, an ensemble Kalman Filter data assimilation approach has been proposed in \cite{talesh2022real}. From a geometry perspective, a geochemistry-based Bayesian deformable surface (GC-BDS) model is described in \cite{leung2021bayesian}. This technique, also known as surface warping, rectifies inaccuracies in existing boundaries by finding the required displacement that minimizes discrepancies between predicted and observed grades. It was subsequently extended using ML with the objective of estimating the domain likelihood for a given displacement, irrespective of the ore-type or geochemical composition \cite{leung2022warpml}.

Predicting the geometric trend (e.g. bedding angle of rock layers in a sedimentary sequence) below the operating bench has been a long-standing challenge. Research on directional prediction is limited, however there is some evidence to suggest that sequence analysis \cite{george2021bedding} and the use of partial differential equations\,/\,variational models may be viable (see supplementary material in \cite{leung2019subsurface}). Model validation is one area that lacks consistency. To address this, concise error metrics and systematic approaches have been suggested in \cite{leung2021bayesian} to facilitate model performance comparison at a glance. These methods may be used in conjunction with compositional data analysis \cite{leung2019sample} to improve resilience against outliers and incorporate geochemical variability \cite{leung2022msfgc}.

\subsection*{Geotechnical engineering, pit development}
At its core, geotechnical engineering in an open-pit mine is about evaluating the stability of individual benches and preventing catastrophic failure, viz. landslips along the slope. Since excavation and disposal of material both add to the cost of mining and carbon footprint, operators would seek to make the slopes of the pit as steep as possible to reduce the volume that needs to be excavated. This objective has to be balanced against the elevated risk of slope failure as the pit gets steeper, which can result in major disruption to mining operations and even the loss of life. Decision theoretic frameworks are used to manage slope stability risk to ensure safe and economic operation. Systematic risk assessment requires estimating the properties of materials in the slopes, understanding the structural weaknesses in the mine, calculating stability for various slope geometries, and monitoring slope performance as the pit is developed, whereas risk management involves making decisions and taking countermeasures (e.g. using anchors, retaining structures or temporarily suspending operations) to mitigate such risks.\footnote{Risk is defined as $R=p(T)\times\sum_i p(C_i\!\mid\! T) u(C_i)$, where $p(T)$ describes the hazard (probability a threat occurs within a given time period), $p(C_i\!\mid\! T)$ denotes the vulnerability or conditional probability that a consequence (e.g. injury, economic damage) occurs given the threat, and $u(C_i)$ denotes the utility or impact of a consequence.} These ideas are further described in \cite{karam2016slope} and \cite{read2009guidelines}.

Slope stability refers to the ability of an inclined slope to withstand its own weight and external forces without experiencing displacement. It depends on intrinsic properties such as rock strength, the existence and persistence of fractures and joint sets, as well as external factors such as blasting, rainfall intensity/duration, infiltration and surface runoff. While the mine geology is usually well-known, knowledge of the geotechnical properties of soils and rocks is often limited---especially the cohesion (shear strength) to resist gravitational load---and these parameters can vary significantly in a small area. Hence, risk management necessarily involves making decisions with incomplete information. This is further complicated by water table and subsurface flow regimes (hydraulic conditions) since the percolation of pore fluids can have a major effect on stability. Traditionally, site characterization requires interpretations of regional geology and inductive reasoning which combines topographic maps, borehole logs and other data with visual observations. This process is generally not as detailed as engineers would like as exploration activities are constrained by time and cost. Deductive reasoning, on the other hand, can supplement this with continuous measurements and time series analytic capabilities. In \cite{karam2016slope}, Karam et al. described a successful application of an early warning system that uses a sensor network which included strain gauges at the Nanfen open-pit iron ore mine. Using AI algorithms, it is possible to detect signs of weakness and raise an alert before physical movement becomes apparent.\ignore{Other monitoring techniques based on Synthetic Aperture Radar (SAR) and near-surface geophysics were also considered.}

In geologically complex areas, it is not uncommon to find strata on opposite sides of a fault dipping in different directions and having different geotechnical properties \cite{karam2016slope}. Significant folding and local deviations in the fault pattern may increase vulnerability and the possibility of wedge failure. Hence, new knowledge obtained from site inspection or targeted drilling should be used to rectify inaccurate assumptions. This can affect how safety is factored into the pit design, how benches and access ramps \cite{kaabachi2021apcom} are constructed. For a concrete illustration, refer to Lucas and de Graaf in \cite{lucas2013iterative}. As new geotechnical investigation is conducted on site, the risk profile is updated. Its importance cannot be overstated as a small angular correction can significantly alter the projected location and determine which bench a fault surface intersects with. Given the stochastic nature of such variability, current research aims to model stratigraphic structures and incrementally update the faults using Bayesian techniques. For instance, a GP may be used to compute the posterior distribution of a fault surface as new wall mappings (exposed surface sections) and geophysical loggings become available. This will progressively shrink the uncertainty envelop and provide a more confident probabilistic risk assessment and robust design for the pit below the current bench.

\begin{figure}[!htb]
\centering
\includegraphics[width=83mm,trim={0mm 0mm 0mm 0mm},clip]{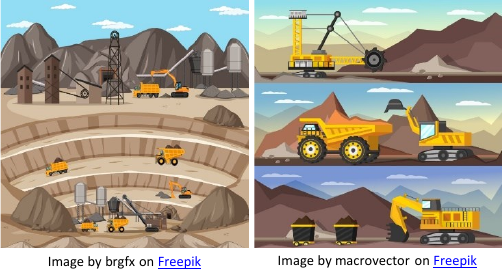}
\caption{Conceptualization of theme 2: pit design (left) and themes 5-6: vehicle dispatch, trains and mine planning (right)}\label{fig:conceptualize-themes}
\end{figure}

\vspace{-3mm} 
\subsection*{Intelligent systems for material tracking and probabilistic inference}
Probabilistic reasoning and big data analytics \cite{choi2022optimization} are essential to many intelligent systems in mining. Although the possibilities are endless, this section will highlight three applications at the frontier of current research. The first example is dynamic road network generation, where vehicle movements are analyzed using haul truck GPS data to produce navigation maps. In \cite{seiler2020haulroad}, Seiler examined different map inference techniques (direct incremental methods, point clustering, kernel density estimation and intersection linking) and made critical observations on their effectiveness in terms of lane separation and attributes specific to mining. This work facilitates other objectives such as multi-vehicle trajectory optimization \cite{gun2019multi}.

The second example is high-resolution material tracking where the dynamic states (e.g. uncertainty and expected grade or volume of material in each voxelated region \cite{innes2011estimation}) of the mine are estimated by propagating beliefs in a probabilistic graphical model. Specifically, changes in material composition are brought about by excavation and transferal of material typically from pits to stockpiles/plants. These constitute local updates which are handled by message passing. In \cite{bailey2022jt}, Bailey considered numerical stability issues and proposed a moment-form junction tree for Gaussian Bayesian networks which target mixture models with semidefinite matrices.

 The final example is activity/material inference using statistical or machine learning approaches to facilitate precise tracking of trucks and ore properties in open-pit mines. In \cite{markham2022load}, Markham et al. proposed a hidden semi-Markov model for load-haul cycle segmentation using vehicle telemetry data. The need for this arises because many mines today still employ legacy systems with limited sensing which contributes to poor quality or missing data. In \cite{kim2020multi}, deep neural networks were used to classify actions using videos captured by cameras mounted on excavators. This is used to monitor the productivity of earthmoving equipment in the construction industry. Although adaptation of similar technology is lagging in the mining industry, its importance is gradually being felt. Research on computer vision and ML for mining is trending up \cite{fu2020deep,valencia2021apcom}. One proposal uses digging equipment as a sensor to distinguish merged material groups based on excavator dynamics \cite{liu2021ajcai}.

\subsection*{Robotics: sensing and automation}
\begin{figure*}[!tbh]
\centering
\includegraphics[width=\textwidth,trim={0mm 0mm 0mm 0mm},clip]{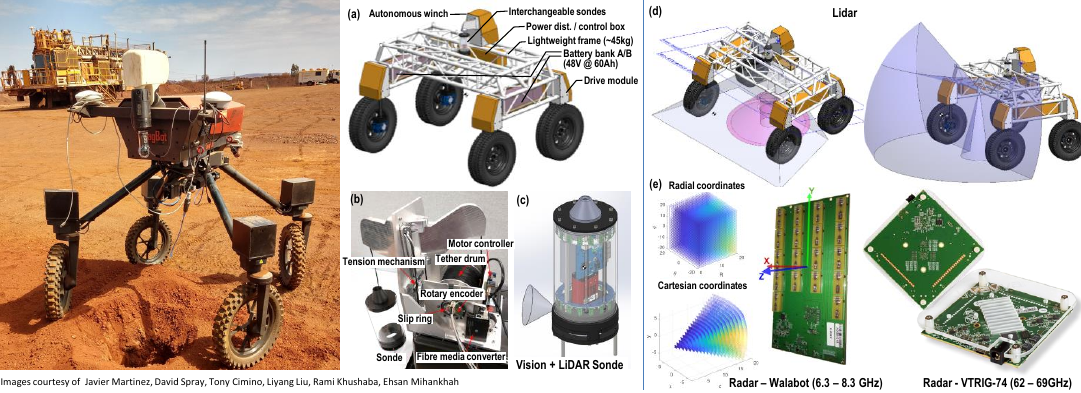}
\caption{A mining-oriented blast-hole inspection robot. (a) System assembly, (b) mechanical parts, (c) vision+LiDAR sonde, (d) LiDAR, (e) radar hardware.} \label{fig:bhir}
\end{figure*}

Robotics can fundamentally improve safety and efficiency in well structured mining problems. For instance, the blasthole DHIR \cite{liu2022acra} and conveyor belt inspection robots \cite{stefaniak2021apcom} can excel in segmentation and localization tasks where the background and geometry remain fairly consistent \cite{zhang2022semi}. In open-pit mines, safety hazards include crush injuries, land slip due to slope failure, and asbestos inhalation from fibrous material. Exposure to these risks can be minimized by assigning detection, inspection or containment tasks to autonomous agents. Beyond collision avoidance and operational data monitoring, which are now common features in mining equipment, appropriate applications of sensing technologies and intelligent systems will be crucial in meeting future needs. For orebody modeling, data fusion utilizes both sparse, precise geochemical measurements from exploration drillings, as well as dense, low-resolution, incidental assay data from production drillings. This static regime leaves big knowledge gaps at unsampled locations. With a mobile platform, adaptive sampling \cite{ahsan-15} using reinforcement learning \cite{levinson2021apcom} or compressed sensing \cite{candes2011probabilistic,calderon2015reconstruction} can provide a cost effective and coordinated strategy for information gathering, where purposeful actions are taken, perhaps cooperatively, to reduce the epistemic uncertainty as models evolve \cite{kumral2013planning}. Depending on the characteristics of the orebody, different sensing modalities may be employed by the robot.\footnote{Established sensing techniques include ground penetrating radar (for bauxite) \cite{ball2021geologist}, bulk density/gamma radiometry (for shale bands) \cite{silversides2017identification}, geomechanical force sensors (for coal) \cite{leung2015automated}, and real-time elemental analysis using a \href{https://sodern.com/wp-content/uploads/2021/12/FastGrade-v6.pdf}{pulsed neutron sensor} in Cu/Fe deposits.}

OEM innovations also help push boundaries. For instance, the ShovelSense technology \cite{haest2022shovelsense} integrates high-speed X-ray fluorescence sensors with excavators to facilitate grade measurements (in-pit material characterization) at unprecedented resolution. In mining, dust, heat, abrasion and vibration are some of the environmental factors that must be considered in robust robotic designs. The fact that different categories of material can have similar chemistry and that ochreous goethite and shale are often confused due to color similarity poses specific challenges for perception/classification. Research in hyperspectral imaging \cite{murphy2012evaluating,murphy2013mapping} has provided ways for resolving these differences and handling intra-class variability. However, deployment has been limited due to cost and other obstacles. One practical issue is that shadows are cast on rough surfaces on mine faces at different times of day. Illumination changes and moving shadows can confound and degrade the performance of hyperspectral mineral classifiers. This effectively reduces the operating window as training data are often obtained under specific lighting conditions. Recent works have focused on robust and generalizable approaches. For instance, Windrim et al. proposed a self-supervised, shadow-invariant DNN architecture using stacked auto-encoders, CNN with lighting augmentation and transfer learning to overcome these difficulties \cite{windrim2023gsf}.

\textbf{Down-hole inspection robot (DHIR)}\label{sec:visualisation-dhir}---A practical example of a novel robotic solution to safety and efficiency is shown in Fig.~\ref{fig:bhir}; the DHIR platform is currently under development. Fig.~\ref{fig:bhir}(left panel) shows an experimental blasthole inspection robot being tested at a mine site. The main purposes of the DHIR are to minimize manual handling of blasthole measurements, and enable real-time data collection to inform blasthole design and charging processes (which correspond to steps \circled{4} and \circled{5} in Fig.~\ref{fig:iron-ore-open-pit-mining-pilbara}). A key motivation behind its design is to overcome the lack of feedback between the planning and drilling stage, this is where automated verification on drilling accuracy as part of QA/QC could improve conformance, and therefore blast fragmentation outcomes. Another reason is to study the structural integrity of drilled holes over time (in between drilling and charging) as hole collapsing events, cavities formation and weathering can influence the effectiveness of a drill and blast strategy.

The DHIR comprises a suite of visual, radar and LiDAR sensors \cite{liu2022acra}. Additional down-hole sensing payloads may be deployed via the winching mechanism depicted in Fig.~\ref{fig:bhir}(a)-(b). The sonde (see Fig.~\ref{fig:bhir}c) is equipped with 32 LiDAR sensors (for measuring the drilled hole diameter/volume), forward-facing RGB camera, dimmable LED light rings, float-based water detection, infrared thermometer (for wall surface temperature measurements), and gigabit communications via the tether. Fig.~\ref{fig:bhir}(d) shows two Ouster LiDAR sensors which are used to detect and straddle blast-cones, and for precise positioning of the sonde above the blast-hole. The radar systems shown in Fig.~\ref{fig:bhir}(e) may be used for object detection, material identification\cite{khushaba2022radar}, gesture recognition, to analyze human-machine interactions or enforce safety in the context of a mine.

\setcounter{table}{0}
\begin{table*}[t]
\setlength{\tabcolsep}{4pt}
\centering
\resizebox{\textwidth}{!}{
\begin{tabular}{|lp{35mm}|p{70mm}|p{92mm}|}\hline
& Mining themes & Disciplines\,/\,primary concerns & Techniques\,/\,general approaches\\\hline
5 & Vehicle dispatch, train timetables & Scheduling. Flow planning. System dynamics. Model predictive control. Collaborative optimization. Transition to renewable. & Integrated control hierarchy. Monte Carlo Tree Search (MCTS). Transition graphs. Hybrid simulation. MILP\\\hline
6 & Mine planning, production scheduling & Operations research. Stochastic optimization. Energy-aware planning algorithms. Integration challenge encompassing energy network. & Discrete event simulation. Agent-based modeling. Scenario planning. Digital twins. Distributed particle simulation. High performance computing. Virtual reality.\\\hline
7 & Optimization of operations for energy efficient, carbon-neutral mining & Sustainability and decarbonization. Zero carbon emissions (targets+pathways). & Eliminate or significantly reduce dependence on diesel and non-renewable electricity generation. Hydrogen fuel-cells and battery electric vehicles. Adapt mining operations and energy network to cope with fluctuating supply and demand\\\hline
8 & Mine remediation/land rehabilitation & Environmental and ecological surveys. Ecosystems and biodiversity. Biotechnology. Remote sensing. Robotic sampling/inspection.	& Life-of-mine management plan. Policy to preserve and restore habitats. Consultation with stakeholders. Local knowledge. Phytoremediation. Meta-genomic survey of soil microbes. UAV. Longitudinal studies\\\hline
\end{tabular}}
\caption{A quick reference for discussion: Themes--technologies matrix for surface mining in the Pilbara (continues from p.\,\pageref{tab:themes-technologies})}\label{tab:themes-technologies-continued}
\end{table*}

As mining robotic technologies continue to evolve, persistent autonomy and interoperability will be central pillars in future research. The former raises issues of safety, security, trust \cite{brundage2020toward} and oversight with humans in-the-loop. The latter is especially relevant as hybrid technologies and novel robotic platforms like DHIR proliferate---increasing reliability in a system of systems, produced by competing vendors, and utilizing heterogeneous data from a distributed system to accomplish higher goals will be significant research topics. Compatibility issues may be addressed, at least in part, by developing open-source software, architectures \cite{pyda2021apcom} and protocols through standardization; with emphasis on machine interactions, communication, data quality and backend integration specific to mining. Integrating AI/ML capabilities into software and services in a production setting poses its own unique challenges. Some real-world lessons and best practices for testing and reducing maintenance costs (technical debt) are presented in \cite{sculley2015hidden,amershi2019software,breck2017ml}.

\subsection*{Vehicle dispatch/train timetables}
As the mining industry's adoption of robotics, automation technology and digital systems grows, there are increasing opportunities to embed increasingly intelligent or optimal controllers such as fleet management systems and scheduling tools in its operations. Deployment of effective control systems typically requires reliable digital communications, a mathematical model and state tracking of the system being controlled (see digital twin \cite{servin2021digital}), a reliable response from the system to control commands, and a clearly defined objective. These components are typically required or provided by the deployment of robotic platforms and automated processes, so there is a significant opportunity to extract increased efficiency through optimized planning and control, in addition to any gains from the direct benefits of automation and robotics. By building an increasingly intelligent and integrated control hierarchy, operations can be made more efficient and reactive to change or new information.

In \cite{gun2019multi}, an online haul-truck velocity optimizer is presented, in which truck speeds are collaboratively optimized to maximize haulage productivity, reducing unnecessary stops at intersections for traffic and queuing \cite{putri2021apcom} at dig and dump points. The feasibility of such a system is dependent on real-time communication as well as an interface to command truck speeds, with guarantees on commands being followed. The coordination of the fleet is an additional advantage, enabled by the automation of trucks, which can respond to regularly changing speed profiles, and reliably drive in a way that may appear (to a human driver) unnecessarily slow, in order to arrive just in time for loading.

Optimization at the fleet level is also demonstrated through advances in dispatch algorithms for fleet management systems, as in \cite{seiler2020flow,samavati2019improvements,carvalho2021apcom}. The application of modern optimization techniques and re-thinking of performance objectives enabled productivity improvements over the algorithm still widely used in commercial fleet management systems \cite{white1993improving,white1986computer}. The new approach taken in formulating dispatch optimization problems has been to optimize for productivity over time, informed by a system model, rather than using heuristic proxies for productivity, like assigning trucks to dig units with short queues or a lack of recent service by trucks. This allows an automated control loop around production targets, rather than using an operator to provide the control feedback, including high-frequency decision making and re-assignment, without the frustration of constant interaction and re-tasking for humans.

Similar advances have been made across the wider supply chain, with modern freight planning on a rail network \cite{vujanic2021computationally,jones2022trainjournal}. There are many tools for static timetable optimisation \cite{shakibayifar2017stochastic,zhang2021simultaneous}, online rescheduling \cite{liao2020real}, conflict detection and resolution\footnote{This means taking an existing timetable and fixing it when some delay creates a conflict in the schedule, rather than generating it periodically from scratch.} \cite{dariano2008reordering}, which are effective for commuter networks or those interacting with other organizations in the supply chain requiring certainty about departures and arrivals. For a more vertically integrated supply chain such as Rio Tinto's iron ore operations in the Pilbara, or well-integrated supply chains of the future, more flexible and efficient operation is achieved without a fixed timetable. The trains that deliver ore product to the ports are largely interchangeable, and many of the ore products are also interchangeable, subject to simple resource constraints. This creates a very large search space for the problem of generating train destinations and motion schedules. \citeauthor{vujanic2021computationally} demonstrate a novel, deadlock-free train scheduler, with run-times in the order of seconds to minutes, compared to the hours it takes to regenerate timetables manually. This facilitates rapid response to incidents, as well as a control loop for integrated, automated train control. Again, this rapid re-scheduling \cite{vujanic2021computationally} and even destination swapping \cite{jones2022trainjournal} is ideally applied to autonomous trains, such as Rio Tinto's AutoHaul System \cite{rio-autohaul2019}, where train meets and stops, destinations and efficient speed commands can be updated in real-time as events unfold, and new information becomes available across the network for globally optimal decision-making.

The future of ore transport is controlled by tightly-integrated, model-predictive controllers, often making heavy use of optimization techniques to model and optimize the effects of many real-world decisions. Integration across hierarchical silos is vital for real-time decision-making that avoids myopic, locally optimal decisions, but instead seeks out the best behavior across the full supply chain.

\vspace{-3mm}
\subsection*{Mine planning and production scheduling}
The scope of mine planning and production scheduling is extensive and usually goes beyond material transportation and network routing; the main ideas featured in the previous section. The scope of such problems began in the 1960's with parametric models focused on the geometric progression of a pit \cite{lerchs1965optimum}, and has grown into a large field of research, with much wider considerations. Optimization methods for mine planning typically involved deterministic orebody models and algorithms, and a staged solution approach, dividing the problem into determining pit limits, mine block geometry, and block sequencing, for example.

Mine planning has evolved to use more integrated methods \cite{osanloo2008long}, though as Osanloo et al. note these rapidly become infeasible with exact methods. Additionally, there has been recognition that maximizing value requires consideration of uncertainties in the orebody, production and even external economic factors \cite{newman2010review}. Modern planning increasingly involves producer-consumer considerations, dynamic systems \cite{uludag2021apcom}, goal setting, resource allocation and combinatorial problems, and the scope and scale of mine planning problems is increasing in line with available computational performance. The objectives and optimization horizons vary depending on how the problem is formulated, and operations are typically driven by a hierarchy of optimizations of increasing horizon and decreasing detail.

As general computational performance has increased, short-term mine planning (horizons less than 1-2 years) has become a larger focus for operations research, and commercial mine planning software \cite{blom2019short,otto2021apcom}. Short-term mine planning features higher fidelity models of the ore movement, stockpiling, rehandling and equipment utilization. In both the academic and commercial worlds, we see an increasing use of event simulation and stochastic optimization methods, and a diverse range of planning algorithms for different timescales and operational requirements. Production scheduling problems are generally NP-hard \cite{moreno2010knapsack}, and thus meta-heuristics are employed in larger problems. For instance, genetic algorithms \cite{muke2021apcom} and dual interchange algorithms\footnote{Dual Interchange is a parent algorithm which supervises two subservient algorithms: simulated annealing and Tabu search \cite{phillips2021apcom}.} have been used to optimize long-term open-pit mine production schedules. Also, hybrid meta-heuristic algorithms that combine GRASP (greedy randomized adaptive search procedure) with mixed integer programming \cite{kenny2017towards}, and Lagrangian relaxation with firefly algorithm \cite{tolouei2021apcom}, have been proposed for precedence constrained production scheduling problems in mining.

A constant challenge is finding feasible and robust solutions that satisfy constraints in a cost effective and resilient way, taking into account the risks posed by a set of probable scenarios. These risks encompass the composition uncertainty of the excavated material, ore blending requirements, capital and operating costs, as well as contractual obligations. We believe that realistic scenarios and parameters are best informed by stochastic models relating to the pits and material movements, supported by reliable and high-fidelity data regularly collected from real-world operations. As models become more complex, digital twins and online tuning of these is required to adapt the models to match the real-world operations. Data collection and integrity must be highly valued by miners.

Expanding on the findings of \cite{blom2019short}, there is an integration challenge in production planning. Not only are there hierarchical planning horizons (long-term and short-term), but further up the supply chain, and all the way down to vehicle control, we see a grand hierarchy reaching from motion control through to economic and geopolitical modeling. Tactical and short-term planning (weeks to months) is further complicated as it becomes difficult to maintain a neat hierarchy of control. Scheduling parallel operations within the pit (extraction, maintenance, road-building), and even different operational modes within extraction, will not always fit into a simple hierarchy.

The transition to renewable energy will further complicate this integration challenge, as energy generation becomes more variable, subject to weather, seasonal and climate effects over various timescales. Energy storage is expensive, and a trade-off will be necessary between capital investment in storage and operational consistency. All levels of production planning will be impacted by energy availability, and the cost of storage to firm across dips in generation. New algorithms will be required for optimization of power storage and distribution, charging mobile equipment, and energy-conscious mine plans, which account for climate and seasonal impacts as well as providing contingencies for still and cloudy days or weeks. These emerging trends mean production planning research will increasingly need to focus on more scalable stochastic algorithms, stochastic and high-fidelity digital-twin-style models, novel methods of coupling or integrating planners, and energy-aware planning at all time-scales.

\vspace{-3mm} 
\subsection*{Optimization of operations for energy\\efficient carbon-neutral mining}
Sustainability and decarbonization are high on the agenda of the mining industry as it faces increasing pressure from regulators, investors and customers to reduce its carbon footprint. As capital markets embrace green industries, miners with lower ESG (environment, social and governance) scores have to face higher cost to access capital to finance mining projects. On the supplier side, leading automotive OEMs have set ambitions for material decarbonization. On the consumer side, zero-carbon producers are emerging, this will increase demand over time for low-carbon iron ore feed for steel production. The financial calculus, mirroring behavior of competitors, societal expectations and consumer demands are driving changes in the industry, toward a potential zero-carbon mining future. To achieve the 1.5$^\circ$C climate-change target by 2050, portfolio changes alone (e.g. divestment of coal assets) will not be enough. The mining industry will need to reduce direct CO$_2$ emissions to zero.

Large miners like Rio Tinto and BHP have communicated ambitious targets for their CO$_2$ emissions reductions to the market, with reduction targets of 50\% and 30\% (respectively) by 2030 and net-zero by 2050 \cite{rio-annualreport2022,bhp-annualreport2022}. These targets relate to Scope 1 (diesel used in mobile equipment) and Scope 2 (non-renewable electricity generation) which account for 40-50\% and 30-35\% of current CO$_2$ emissions, respectively. In \cite{legge2021zero}, Legge et al. showed emissions intensity varies widely across iron and copper ore mines, and showed haul trucks are the single biggest source of emissions from the mine, accounting for 35\% of the total, while crushers, bulldozers and excavators account for 20\%, 7\% and 5\%, respectively. Together, they contribute to two third of the overall emissions, highlighting the challenge for mining companies to focus on transition of mobile equipment to carbon-neutral energy sources, and a need to optimize the efficiency of this equipment to make that transition affordable. This will require further advances to the equipment and fleet optimization discussed earlier in this section.

The McKinsey report \cite{legge2021zero} canvassed options for decarbonization and considered their cost and complexity. The main ones include the use of sustainable (bio)fuels, shift to hydrogen fuel cells and battery electric vehicles (FCEV and BEV), and supplementing with green electricity (e.g. generated from wind/solar). Large-scale initiatives are already under development at Rio Tinto's Gudai-Darri mine which aims to satisfy 65\% of its total electricity consumption using solar farms and battery storage. These options are becoming viable as the total cost of ownership for BEV/FCEV haul truck comes down. The report concludes that ``decarbonization presents a significant opportunity for ambitious players to differentiate themselves and lead the way toward zero-carbon mining. For this to happen at scale across the industry, multiple stakeholders need to work together to develop the potentially cost-positive abatement approaches that are currently unavailable'' \cite{legge2021zero}.

Engineering is part of the solution. Since renewable energy sources are largely stochastic, time and weather dependent, one foreseeable challenge is adapting mining and energy network operations to deal with fluctuating supply and demand. Short-term mine planning may strive to balance efficiency with redundancy (for instance, with reasonable extra capacity and a view of future redeployment) to keep the flow rate constant by maintaining some sort of buffer. Long-term optimization requires robust strategies that consider how supply (fleet, grade, volume) and blending constraints can be satisfied at reasonable cost, and deal with disruptions and uncertainties. Future work in process control, mine planning and industrial processes generally will need to consider the possibility of more dynamic daily and seasonal optimization. Learning from the agriculture industry, mining will likely see a transition from 24/7 operation to `digging ore while the sun shines'.

A further insight is that if the highest emitting quartile of mining operations were brought in line with industry average, total emissions would be reduced by $\sim$46\% (34Mt) for iron ore and $\sim$26\% (19Mt) for copper. From the report, we see these higher emitters are generally smaller operations (by percentage of global production), highlighting the need for democratization of technological advances supporting carbon emission reduction. Advances in efficiency and algorithms to make the best use of energy storage or support effective recharging schedules, for example, should be made available to all mining companies and vendors to facilitate a rapid reduction in global emissions.

\subsection*{Mine remediation/land rehabilitation}
\begin{table*}[!tbh]
\setlength{\tabcolsep}{4pt}
\centering
\resizebox{\textwidth}{!}{
\begin{tabular}{|lp{60mm}|p{160mm}|}\hline %45mm,120mm
& Mining themes & Key challenges and possible future work\\\hline
1 & Geology modeling &
Rapid incremental geological boundary update and data assimilation approaches. Obtaining reliable (heteroscedastic aleatoric) uncertainty and mean grade estimates using production data such as blasthole geochemical assays. Increasing the density and consistency of usable data through large-scale autonomous geological data collection.\\\hline
2 & Pit development and slope stability &
Probabilistic risk assessment with incomplete information. Integrating expert knowledge, logs and maps from geotechnical surveys, and heterogeneous data collected from sensor networks, satellites and field robotic platforms to detect structural vulnerabilities and prevent slope failure. Quantifying the value of additional data to optimally direct campaigns to fill in structural knowledge gap.\\\hline
3 & Intelligent systems for material tracking and probabilistic inference &
Using telemetry along-side emerging perception (computer vision) and probabilistic modeling techniques to build an accurate digital twin for the mine, that quantifies its own uncertainty. Accurately account for all material movement across load, dump and re-handling activities. Update grade control and other planning tools to make use of stochastic models of mine state and material movement.\\\hline
4 & Robotics: sensing and automation &
Interoperability, persistent autonomy, reliability engineering of complex systems with humans-in-the-loop. Standardization, co-design and development of protocols, covering aspects such as actuation, peer-to-peer/wireless communication, quality assurance, and back-end integration specific to mining. There is scope for interdisciplinary research, technology showcase and industry engagement through TEDx/expo/symposium, targeting domain-specific robotic challenges. Topics of interest include field robotics for surface mining, robust design, adaptive sampling, reinforcement learning, collaborative information gathering, novel sensing technologies and hardware-software integration.\\\hline
5 & Vehicle dispatch, train timetables &
Implement and validate scheduling algorithms, with highly accurate digital twins and simulations. Valuing the data required to create digital twins and simulations enough to collect more than the bare minimum for operations. Re-evaluating changing control modes (strategies) under different circumstances, without being confined to a rigid control algorithm or hierarchy. Data integrity issues and unobserved system states. Ability to re-plan in real-time following service disruptions or unexpected events. Transition to renewable energy sources will bring significant new constraints to consider in control algorithms.\\\hline
6 & Mine planning, production scheduling &
Stochastic optimization and scenario planning need to factor in energy constraints (daily and seasonal cycles, weather conditions and events). Increasing demand for high performance computing and simulation to distill questions, validate approaches and instill confidence in stakeholders---similar to interpretable machine learning. Better coupling of planning layers (life-of-mine to daily plans) is computationally challenging, but will improve planning outcomes at all levels.\\\hline
7 & Optimization of operations for energy efficient mining &
Significantly reduce dependence on non-renewable electricity generation in the most cost effective way. Adapt mining operations and energy network to cope with fluctuating supply and demand, through a combination of efficiency improvements, storage/firming and changing operational modes to adapt to (cyclic and event-based) energy constraints. Increased demand for critical minerals.\\\hline
8 & Mine remediation / land rehabilitation &
Sustained environmental and ecological recovery post-mining. Manage biodiversity and ecosystem using scalable and coordinated strategies. These may include phytoremediation, biotechnology, remote sensing, robotic platforms for sampling and cultivation, and artificial intelligence to guide these processes.\\\hline
\multicolumn{3}{c}{}\\
\end{tabular}}
\caption{Challenges and possible future work across the mining themes}\label{tab:challenges}
\end{table*}
Land rehabilitation post-mining is another issue of immense importance. To earn a social license to operate, consultation with traditional land owners and a comprehensive mine rehabilitation strategy should be integral parts in the life-of-mine management plan. Some guiding principles and experiences are described in \cite{bell2001establishment,erickson2017benefits,grant2016rehab}. An essential feature is a shift in focus from extraction to a commitment to preserve and restore habitats post-mining from the beginning. This might entail environmental/ecological surveys, collection of seeds for native species or relocation of vegetation cover to a temporary site before excavation begins. A well thought-out plan should encompass multidisciplinary perspectives, which typically includes an assessment of biodiversity, establishing targets and pathways for the restoration of ecosystems. In addition, biotechnology and biochemical engineering of mine wastes also play a major role in sustainable recovery and pollution control \cite{uq-smi2023}.

In the Australian context, research had found ``the use of native plant seeds is fundamental to large-scale rehabilitation and the re-establishment of self-sustaining ecosystems after high-impact mining activity'' \cite{erickson2017benefits}. A variety of botanical, ecological and environmental insights---from smoke-derived germination stimulation\cite{erickson2017benefits} to phytoremediation of spoil dumps\cite{banerjee2019vetiver}---offer potential pathways for remediation and restoration. As observed in the ag-robotics industry, the immense scale of mining rehabilitation will benefit from robotic solution to vegetation management. Countries throughout the world have increasingly focused on mining rehabilitation efforts and many have reported promising results based on field observations \cite{ranjan2016reclamation}, soil health indicators \cite{rodriguez2021native} and meta-genomic survey of soil microbes \cite{gastauer2019metagenomic}. Remote sensing (via satellite imaging and drones) will likely play a pivotal role in gauging the success of large-scale sustained recovery, particularly in detectinig changes in vegetation. At the macroscopic scale, recent advances in agricultural robotics, computer vision and machine learning \cite{chlingaryan2018machine,pretto2020building} offer agile and scalable solutions for soil health monitoring, sampling, analysis and weed control. This will likely generate large-scale longitudinal data that inform and improve the efficacy and sustainability of mine rehabilitation and land management practices, including soil carbon sequestration \cite{robson2021soilcarbon}.

To recapitulate, we refer readers to Table~\ref{tab:challenges} for a summary of the challenges across the eight mining themes.

\vspace{-3mm}
\section*{\color{CadmiumRed}Concluding remarks}\label{sec:conclusion}
Looking at the diverse and complex technologies used in mining automation, it is perhaps not surprising that its evolution cannot be traced back to a single origin or moment in time\ignore{ascribed to a single cause}. Rather, it comprises parallel streams of progress that seem destined to converge with increasing computational intelligence. One possible development is a gradual deployment of collaborative information gathering semi-autonomous agents---perhaps with the use of novel/scalable sensing techniques in distributed systems and probabilistic fusion of data from heterogeneous sources---to achieve higher and more ambitious goals. These goals may relate to holistic mine planning and dynamic resource allocation or implementing adaptive solutions for energy efficient open-pit operations. An associated challenge is recognizing fundamental shifts in modeled processes and assumptions, and responding robustly to data\,/\,concept drifts and disruptions in a changing environment to enable continuous operations and persistent autonomy within an interconnected system. This observation probably applies to a multitude of problems in perception, geostatistics, machine learning and stochastic optimization within mining.

There are new opportunities for automation across the spectrum in most facets of open-pit mining. Some require specialized knowledge that can be developed through close collaboration with subject matter experts and industry. Others require ingenuity and engineering skill sets that are prevalent in the robotics and AI community. The authors believe significant progress may come from university-industry engagement and meaningful interactions between both groups.

A successful partnership requires patience and goodwill, alignment of business and technical objectives and pursuing common interests under a shared vision. This also entails a willingness to navigate change and be open to learning and using new technologies to achieve strategic goals. A commitment to research and development is arguably a prudent investment decision. It can bring about transformational change, mitigate risks and improve efficiency/productivity in the long-term.

Electrification of mining vehicles--- a reduction in reliance on fossil fuels---currently represents one of the top strategic priorities for many mining companies. As good corporate citizens, there are both economic and environmental imperatives for reducing carbon emissions. With this, comes also the challenge of battery storage and operating a sustainable energy grid to support mining activities.

The experience gained from critical analysis, technology deployment and diversification allows industries to stay ahead of the game and adapt more quickly to future challenges. We hope this article can stimulate interest in the broader community and provide a useful starting point for understanding the problems and opportunities in mining automation.

\vspace{3mm}
\noindent{\Large\textbf{Supplementary Material}}

\vspace{3mm}
\noindent Refer to conference poster and slides on page~{\color{blue}\pageref{fig:mining-automation-vision}}.

\vspace{-3mm}
\section*{\color{AmericanRose}Acknowledgments}
This work has been supported by the Australian Centre for Field Robotics and the Rio Tinto Centre for Mine Automation. The authors would like to acknowledge all students and staff from the past and present who have contributed to the success of RTCMA.

%Next two lines manipulate spacing to achieve an even spread between columns
%\addtolength{\textheight}{-13cm}
%\addtolength{\footskip}{13cm}
%
\renewcommand*{\bibfont}{\footnotesize}%\footnotesize
\setlength{\bibitemsep}{0.1\itemsep}%\itemsep
\vspace{-3mm}
\printbibliography[title={\color{CadmiumRed}References}]
%If main.bbl is deleted, run pdfLaTeX+MakeIndex+BibTeX to generate .bcf file.
%Then, run Biber (e.g. in TeXworks) to generate the .bbl file.

\newpage\clearpage
\section*{\color{CadmiumRed}Picture credits}\label{picture-credits}
\begin{table}[!h]
\vspace*{-\baselineskip}
\includegraphics[]{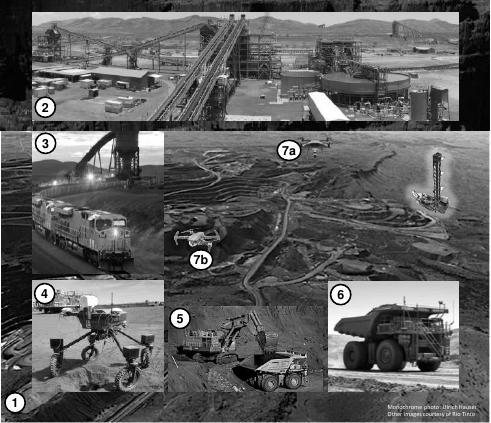}
\vspace*{-\baselineskip}
\end{table}

\begin{table}[!h]
\centering
\resizebox{\columnwidth}{!}{%
\begin{tabular}{l|p{100mm}}
1 & \textit{Tom Price aerial photos}. \textcopyright Ulrich (Ueli) Hauser, Boulder, \nobreak{Colorado}. Used with permission.\\
2 & \textit{Silvergrass project}. Rio Tinto Investor Seminar 2016 - ``Capital Markets Day'', Sydney (\link{\href{https://www.asx.com.au/asxpdf/20161124/pdf/43d504sfq7mv4v.pdf\#page=40}{slide 38}}). \textcopyright Rio Tinto. Release date: 24 November, 2016.\\
3 & \textit{Rio Tinto freight train}. Rio Tinto Investor Seminar, Sydney 2017 (\link{\href{https://www.riotinto.com/-/media/Content/Documents/Invest/Presentations/2017/RT-Investor-Seminar-2017-slides.pdf\#page=1}{slide 1}}). \textcopyright Rio Tinto. Release date: 4 December, 2017.\\%https://www.amsj.com.au/mining-giants-order-eco-friendly-trains/
4 & \textit{Down-hole inspection robot}. \textcopyright Australian Centre for Field Robotics, The University of Sydney.\\
5 & \textit{Excavator operations}. Rio Tinto Investor Seminar 2016 - ``Capital Markets Day'', Sydney (\link{\href{https://www.asx.com.au/asxpdf/20161124/pdf/43d504sfq7mv4v.pdf\#page=15}{slide 13}}). \textcopyright Rio Tinto. Release date: 24 November, 2016.\\
6 & \textit{Haul truck}. Rio Tinto 2021 Half Year Results (\link{\href{https://www.riotinto.com/-/media/Content/Documents/Invest/Financial-news-and-performance/Results/RT-Half-year-results-2021-combined.pdf\#page=12}{slide 12}}). \textcopyright Rio Tinto. Release date: 28 July, 2021.\\
7 & \textit{Drone PNG Transparent Images}. \textcopyright \link{\href{https://www.pngall.com/drone-png}{PNG ALL}}. CC4.0 BY-NC license
\end{tabular}
}
\end{table}

\vspace{-5mm}
\begin{table}[!h]
\includegraphics[]{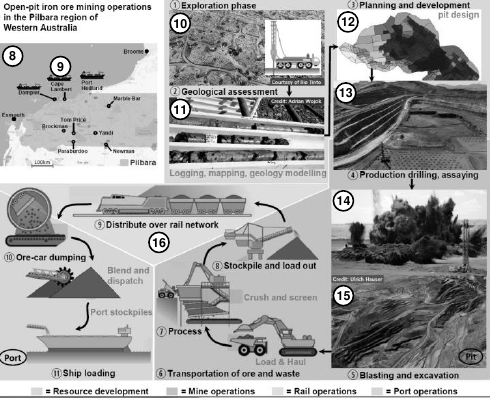}
\end{table}

\begin{table}[!h]
\centering
\resizebox{\columnwidth}{!}{%
\begin{tabular}{l|p{100mm}}
8 & \textit{Pilbara region from Google Map}. \link{\href{https://www.google.com/maps/place/Pilbara,+WA/@-19.8383383,116.9018258,7.5z/data=!4m5!3m4!1s0x2b8c2cc3bfc08385:0xfde4fcf03ec7b5dd!8m2!3d-21.5921433!4d121.5236826}{Map data}} \textcopyright 2023 Google.\\%Under fair use provision
9 & \textit{Ship symbol}. \link{\href{https://www.pngwing.com/en/free-png-nlxul}{Car Ship Watercraft Vehicle}}. Non-commercial use.\\
10 & \textit{Hope Downs iron ore mine}. Courtesy of \link{\href{https://www.mining-technology.com/projects/hope-downs/}{Rio Tinto}}.\\
11 & \textit{Diamond Tip Drilled Rock Cores of Iron Ore -- Australia}. Credit: Adrian Wojcik. Licensed by \link{\href{https://www.istockphoto.com/photo/exploration-drilling-gm509080789-45747190}{iStock}}.\\
12 & \textit{Geochemistry visualisation at an anonymous mine site}. \textcopyright Authors.\\
13 & \textit{A Rio Tinto mine in Pilbara region}. Courtesy of \link{\href{https://www.mining-technology.com/news/newsrio-tinto-mining-operations-to-support-1000-construction-jobs-in-western-australia-5782379/}{Rio Tinto}}.\\
14 & \textit{Blasting at Pannawonica Mine}. Held by \link{\href{https://exhibitions.slwa.wa.gov.au/s/mewa/media/301}{State Library of Western Australia}}. Photo contributed by Rio Tinto.\\
15 & \textit{Tom Price aerial photos}. \textcopyright Ulrich (Ueli) Hauser, Boulder, \nobreak{Colorado}. Used with permission.\\
\end{tabular}
}
\end{table}
\vspace{-5mm}
\begin{table}[!h]
\centering
\resizebox{\columnwidth}{!}{%
\begin{tabular}{l|p{100mm}}
16 & \textit{Line art graphics} adapted from \textit{Pilbara mining process diagram}. Courtesy of \link{\href{https://www.riotinto.com/-/media/Content/Documents/Operations/Pilbara/RT-Pilbara-Mining-process-diagram.pdf}{Rio Tinto}}.\\
\end{tabular}
}
\end{table}
\vspace{-5mm}
\begin{table}[!h]
\includegraphics[]{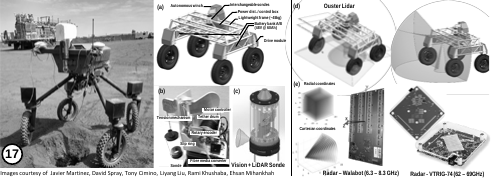}
\end{table}

\vspace{-5mm}
\begin{table}[!h]
\centering
\resizebox{\columnwidth}{!}{%
\begin{tabular}{l|p{100mm}}
17 & \textit{Photos and diagrams relating to mining robotics}. Provided by Javier Martinez et al. \textcopyright Australian Centre for Field Robotics, The University of Sydney.\\
\end{tabular}
}
\end{table}

\vspace{-5mm}
\begin{table}[!h]
\includegraphics[]{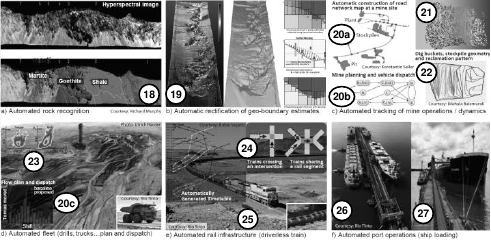}
\end{table}

\vspace{-5mm}
\begin{table}[!h]
\centering
\resizebox{\columnwidth}{!}{%
\begin{tabular}{l|p{100mm}}
18 & \textit{Hyperspectral imaging mineral classification}. Courtesy of Richard J Murphy. \textcopyright IEEE \cite{murphy2012evaluating}\\
19 & \textit{Boundary rectification}. Courtesy of Raymond Leung.\\
20 & \textit{Haul road generation. Online flow planning/dispatch}. Courtesy of Konstantin Seiler. (a)\,\textcopyright SAIMM \cite{seiler2020haulroad}, (b-c)\,\textcopyright IEEE \cite{seiler2020flow}\\
21 & \textit{Dig buckets localization}. Courtesy of Raymond Leung.\\
22 & \textit{Stockpile geometry}. Courtesy of Mehala Balamurali.\\
23 & \textit{Drill monitoring -- geomechanical parameters}. Courtesy of Raymond Leung.\\
24 & \textit{Train scheduling}. Courtesy of Robin Vujanic and William Jones. \textcopyright IEEE \cite{vujanic2021computationally}\\
25 & \textit{AutoHaul rail}. Transportation of iron ore from pit to port (background photo) Courtesy of Rio Tinto.\\
26 & \textit{Rio Tinto port operations}. Rio Tinto Investor Seminar 2016 - ``Capital Markets Day'', Sydney (\link{\href{https://www.asx.com.au/asxpdf/20161124/pdf/43d504sfq7mv4v.pdf\#page=3}{slide 1}}). \textcopyright Rio Tinto. Release date: 24 November 2016\\
27 & \textit{Rio Tinto Weipa Operations}. \textcopyright 2015 \link{\href{https://minedocs.com/17/Weipa_Operations_brochure_2015.pdf}{Rio Tinto}}.\\
\end{tabular}
}
\end{table}

\vspace{-5mm}
\begin{table}[!h]
\includegraphics[]{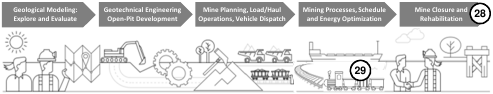}
\end{table}

\vspace{-5mm}
\begin{table}[!h]
\centering
\resizebox{\columnwidth}{!}{%
\begin{tabular}{l|p{100mm}}
28 & \textit{Business model}. Rio Tinto Annual Report 2017 (\link{\href{https://www.riotinto.com/-/media/content/documents/invest/reports/annual-reports/2017/rt-annual-report-2017.pdf\#page=14}{pages 14-15}}). \textcopyright Rio Tinto. Used with modified captions.\\
29 & \textit{Train and track}. \link{\href{https://www.shutterstock.com/search/cartoon-train-black-white}{Train cartoon vector illustration}}. Credit: Nosyrevy. Licensed by \link{\href{https://www.agefotostock.com/age/en/details-photo/coloring-book-train-cartoon-vector-illustration/ESY-043475449}{AgeFotoStock}}.\\
\end{tabular}
}
\end{table}

\vspace{-5mm}
\begin{table}[!h]
\includegraphics[width=60mm]{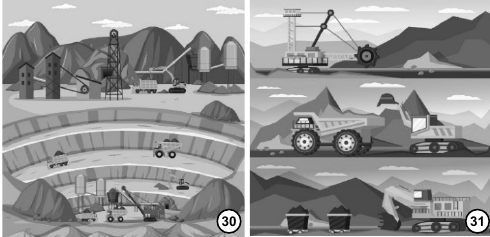}
\end{table}
\begin{table}[!h]
\centering
\resizebox{\columnwidth}{!}{%
\vspace{-5mm}
\begin{tabular}{l|p{100mm}}
30 & \textit{Mining scene with crane trucks}. Image by brgfx on \link{\href{https://www.freepik.com/free-vector/landscape-coal-mining-scene-with-crane-trucks_13642946.htm}{Freepik}}.\\
31 & \textit{Mining industry orthogonal illustrations}. Image by macrovector on \link{\href{https://www.freepik.com/free-vector/mining-industry-orthogonal-illustrationsset_10155393.htm}{Freepik}}.\\
\end{tabular}
}
\end{table}

\vfill

\newpage
\begin{table*}[!h]
\begin{tabular}{p{170mm}}
\textbf{\Large\color{AmericanRose}Supplementary Material}\\
\\
This work was presented at the IEEE International Conference for Robotics \& Automation (ICRA Yokohama) in May 2024. The conference poster and slides are packaged with the \href{https://arxiv.org/src/2301.09771v6}{\color{blue}TeX source} and may be downloaded separately from the following links:\\
\\
\textbullet\, \textbf{ICRA poster} [\href{https://arxiv.org/src/2301.09771v6/anc/icra24-automation-mining-poster.pdf}{\color{blue}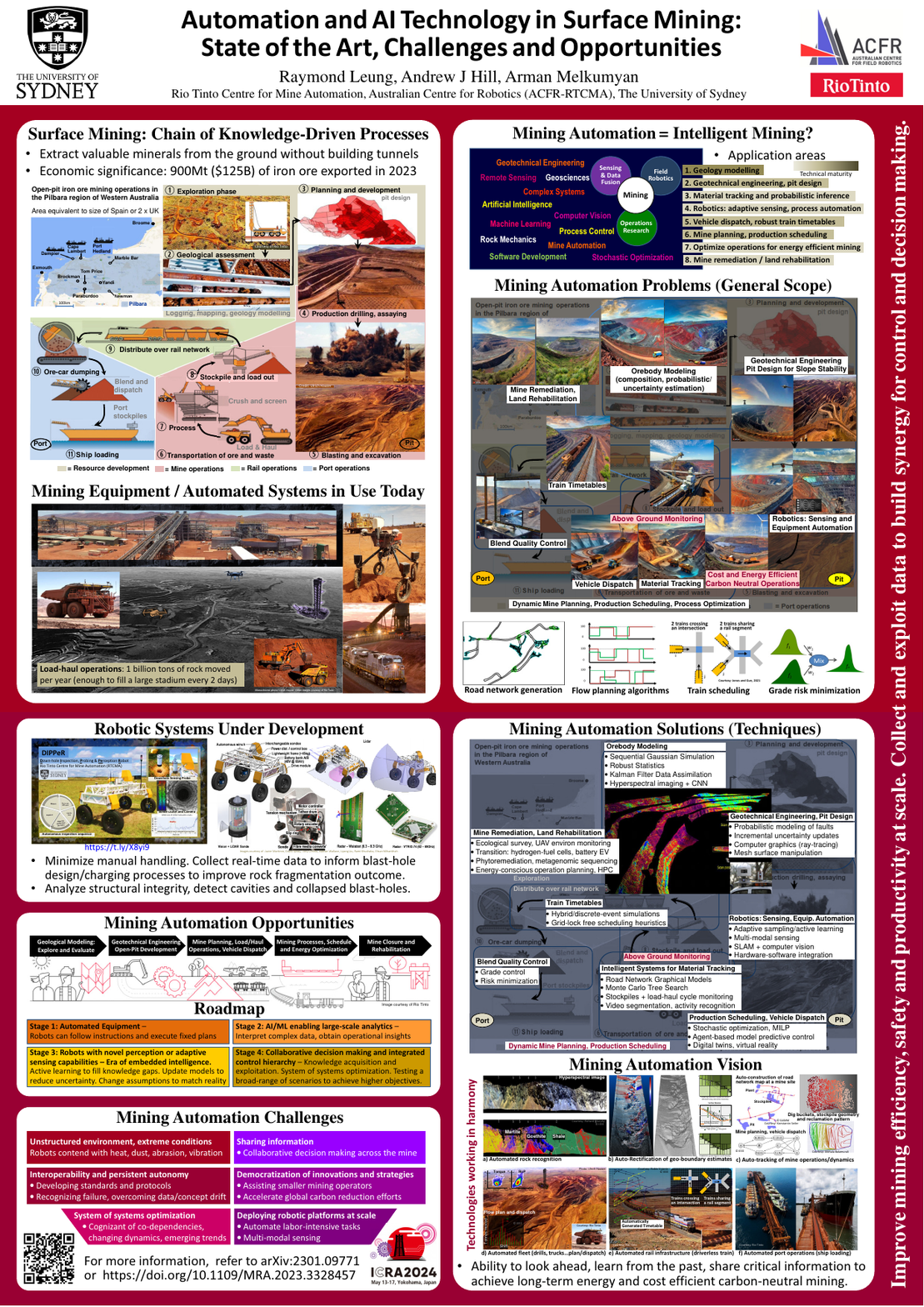}]\\
\textbullet\, \textbf{ICRA annotated slides} [\href{https://arxiv.org/src/2301.09771v6/anc/icra24-automation-mining-slides.pdf}{\color{blue}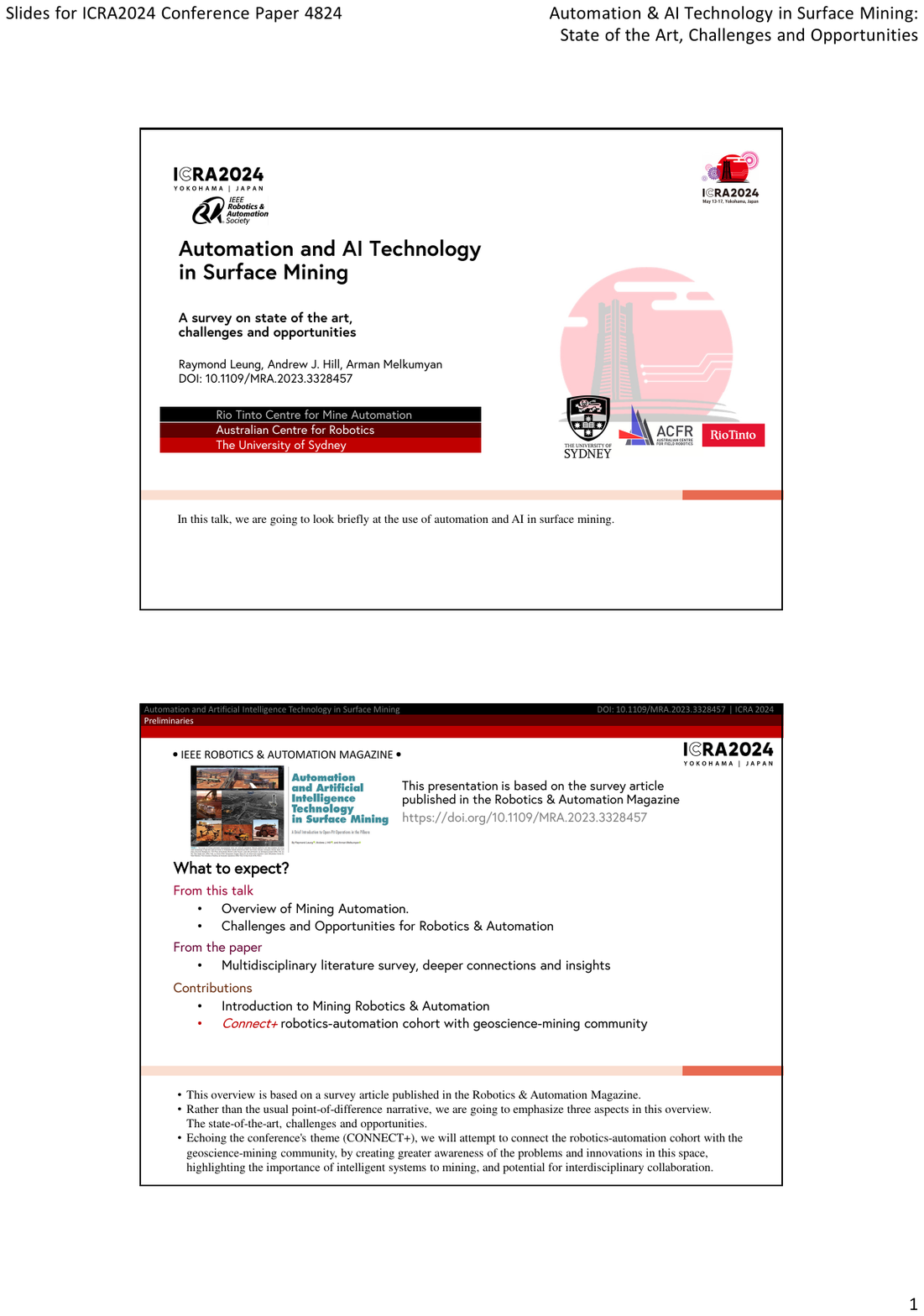}]\\
\\
The following highlights some of the content and discussion.\\
\end{tabular}
\end{table*}

\begin{figure*}[!htb]
\centering
\includegraphics[width=175mm]{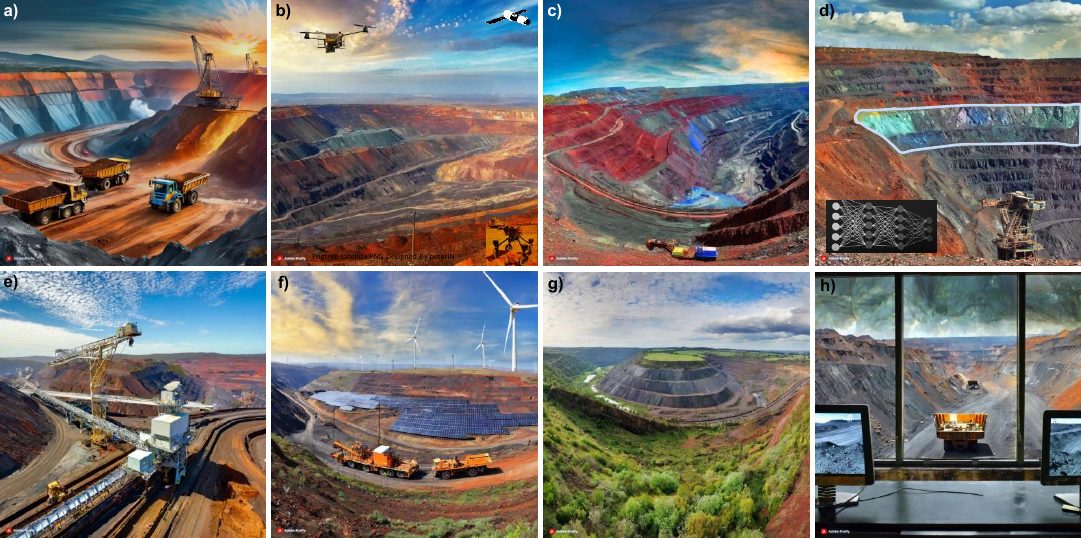}
\caption{The authors' vision for mining automation encompasses (a) robust surface mining operations using automated equipment such as auto-drills, excavators and haul trucks; (b) remote sensing (such as drones and satellite imaging) for inventory management, aerial risk assessment and mine planning, and strategic deployment of ground robots for data collection, site monitoring and inspection across multiple sites; (c) probabilistic interpretation of ore body composition with high-fidelity material tracking and data fusion; (d) new sensing capabilities on mobile platforms (e.g. hyperspectral scans and applications of deep learning); (e) plant optimization and intelligent transportion schedules; (f) cost- and energy efficient carbon-neutral mining, using renewable energy sources (wind and solar), battery-electric vehicles, stochastic/portfolio optimization; (g) mine closure and land rehabilitation considerations being an integral part of long-term mine planning; (h) increased situational awareness facilitated by advanced human/machine interfaces featuring immersive experiences, digital twins, scenario exploration via simulation and whole-of-mine optimization. [Note: these images were created using generative AI.]} \label{fig:mining-automation-vision}
\end{figure*}
\vfill

\end{document}